\documentclass[a4paper,10pt]{article}
\usepackage{latexsym}
\usepackage{amsfonts}
\begin{document}

\newtheorem{definition}{Definition}
\newtheorem{theorem}{Theorem}
\newtheorem{proposition}{Proposition}
\newtheorem{remark}{Remark}
\newtheorem{corollary}{Corollary}
\newtheorem{lemma}{Lemma}
\newtheorem{observation}{Observation}
\newtheorem{fact}{Fact}
\newtheorem{example}{Example}
\newtheorem{conjecture}{Conjecture}

\newcommand{\qed}{\hfill$\Box$\medskip}

\makeatletter
\def\adots{\mathinner{\mkern 1mu\raise \p@ \vbox{\kern 7\p@ \hbox{.}}
\mkern 2mu \raise 4\p@ \hbox{.}\mkern 2mu\raise 7\p@ \hbox{.}\mkern 1mu}}
\makeatother
\newcommand{\ntowers}{NSPACE$\left(2^{2^{\adots^{2}}}\right)$
 }

\title{Efficient Computation by Three Counter Machines} 
\author{Holger Petersen\\
Reinsburgstr.~75\\
70197 Stuttgart\\
Germany} 

\maketitle

\begin{abstract}
We show that multiplication can be done in polynomial time on a three counter machine that receives its input as the contents of 
two counters. The technique is generalized to functions of two variables computable by deterministic Turing machines in linear space.
\end{abstract}

\section{Preliminaries}
In an investigation of the power of simple counter machines \cite{Schroeppel73}, 
Schroeppel described a four counter machine that multiplies two numbers in quadratic time
and posed as a hard problem to multiply two numbers using only three counters. 
He presented a solution based on an exponential encoding of numbers and 
asked whether there is a way that takes less time.

The purpose of this note is to first present a solution for multiplication that runs in time
polynomial in the maximum input value
and then generalize the technique to all  two variable  functions that can be computed by a 
deterministic Turing machines operating in linear space (linear bounded automaton)
to which the arguments are passed in binary representation on its work-tape. This class
of functions includes integer division.

A {\em $k$ counter machine} is equipped with $k$ counters each storing a nonnegative integer. 
Its operation is controlled by a deterministic sequential program consisting of four types of instructions:
\begin{description}
\item[Increment:] Add 1 to the specified counter.
\item[Conditional Decrement:] If  the specified counter stores a value greater than 0, subtract 1. 
Otherwise leave the counter unchanged and jump to a specified instruction out of the normal flow 
of control.
\item[Unconditional Jump:] Flow of control is transferred to the specified instruction.
\item[Halt:] Stop execution.
\end{description}
Notice that unlike many computational models studied in Complexity Theory, a counter machine
has no separate input-tape or output-tape and receives input values as the contents of its counters, 
which corresponds to a unary encoding. 

Instead of  breaking down an algorithm into these four basic instruction, we will use 
additional notation borrowed from higher programming languages. 
In the following we sketch how to simulate these constructs with macros on counter machines.
By R, R1, R2 we denote any of the first $k-1$ counters, while counter $k$ is reserved as 
auxiliary scratch memory. Counter $k$ is assumed to have 0 as its initial value and all macros will
restore this value.
\begin{description}
\item[{\tt if R > 0 then begin\ldots end}:] Decrement R and jump to the instruction after {\tt end} if R 
was 0 before the instruction. 
Otherwise increment R (restoring its value before the {\tt if}) and continue with the instructions 
between {\tt begin} and {\tt end}.
\item[{\tt while R > 0 do begin\ldots end}:] Like {\tt if R > 0 then\ldots} described
above, but jump back to the decrement instruction before the {\tt end}.
\item[{\tt if odd(R) then begin\ldots end}:] In a loop decrement R twice while incrementing counter $k$. 
If R has the value 0 in the first decrement instruction, its inital value was even. Then restore its value by
incrementing R twice while decrementing counter $k$. Skip the {\tt begin \ldots end}
block.
If R has the value 0 in the second decrement instruction of the loop, its inital value was odd.
Restore its value as in the case when its value was even, additionally adding 1. 
Execute the instructions between {\tt begin} and {\tt end},
\item[{\tt while even(R) do begin\ldots end}:] Like {\tt \tt if odd(R) then\ldots} 
described above (roles of odd/even interchanged), but jump back to the test before the {\tt end}.
\item[{\tt R1 := R2}:] In a loop decrement R1 until it is 0.
In another loop decrement R2 while incrementing R1 and counter $k$. Finally in a loop decrement 
counter $k$ and increment R2, restoring its previous value.
\item[{\tt R1 := R1 + R2}:] In a loop decrement R2 until it is 0 while incrementing R1 and counter $k$.
In another loop decrement counter $k$ and increment R2, thus restoring its previous value.
\item[{\tt R1 := R1 - R2}:] In a loop decrement R2 until it is 0 while decrementing R1 (if possible) and 
incrementnig counter $k$.
In another loop decrement counter $k$ and increment R2, thus restoring its previous value. Note
that if R1 $<$ R2, the resulting value of  R1 will be 0. This operation is sometimes called modified minus.
\item[{\tt R := m * R}:]  
In a loop decrement R while incrementing counter $k$. In another loop decrement 
counter $k$ and increment R in every iteration m times. 
\item[{\tt R := R div m}:]  
In a loop decrement R while incrementing counter $k$. In another loop decrement 
counter $k$ m times and increment R for every full iteration. 
\end{description}

\section{Results}
\begin{proposition}
A three counter machine can compute $X*Y$ for nonnegative integers $X$ and $Y$
 in polynomial time as the contents of a counter when
$X$ and $Y$ are initially stored in two counters.
\end{proposition}
{\bf Proof.} The algorithm will be presented using the macros introduced above:

\begin{verbatim} 
procedure mult; (* input: X in A, Y in B; output: B *)
begin
  if B > 0 then (* special case Y = 0, output Y *)
  begin
    A := 2 * A + 1; (* flag in lowest bit *)
(* Loop I *)
    while B > 0 do
    begin
      A := 2 * A;
      if odd(B) then begin A := A + 1 end;
      A := 2 * A;
      B := B div 2
    end;
    B := A;
    A := A + 1;  (* flag in lowest bit *)
(* Loop II *)
    while even(B) do
    begin
      A := 2 * A;
      B := B div 4
    end;
    B := B - 1; (* remove flag  *)
(* Loop III *)
    while even(A) do
    begin
      B := 8 * B;
      A := A div 2
    end;
    A := A - 1;  (* remove flag  *)
(* Loop IV *) 
 while even(A) do 
    begin
      A := A div 2;
      B := B div 4;
      if odd(A) then begin A := A + B end;
      A := A div 2;
      B := B div 2;
    end;
    A := A - B; (* adjust initial value of  A *)
    B := A;
    B := B div 2 (* remove flag  *)
  end
end; (* mult *)
\end{verbatim}
In the following the purpose of the four loops is outlined:
\begin{description}
\item[Loop I:] For each bit $d$ of B (input Y) shift two bits $d0$ into A, thus reversing the order of bits.
\item[Loop II:] Each two-bit group generated in loop I is translated into 0 and shifted into A. 
At the end of loop II counter A contains (starting from lowest bits) $\log_2 (\mbox{Y}+1)$ 0-bits,
$2\log_2 (\mbox{Y}+1)$ groups of two bits each containing a single bit of Y in the higher
order bit, a single 1 and the input X shifted by $3\log_2 (\mbox{Y}+1) + 1$ positions.
At the end of loop II counter B contains $2\mbox{Y}+1$.
\item[Loop III:] B is shifted by $3\log_2 (\mbox{Y}+1)$ positions, while the trailing 0-bits are removed from A.
\item[Loop IV:] Add the multiples of X stored in B to an `accumulator' in A with initial value Y. 
In addition, A stores the two-bit representation of X. It is decoded and controls the additions.
\end{description}
Notice that the maximum value handled by the algorithm is of order $\mbox{X}*\mbox{Y}^3$. The macros
introduced are time-bounded by this value and the loops of the main algorithm are executed 
$\log_2 (\mbox{Y}+1)$ times. Thus the running time of algorithm is polynomial in the input values.
\qed

\begin{theorem}
The class of functions of two variables computable by three counter machines in polynomial time coincides
with the class of  functions of two variables computable by deterministic Turing machines in linear space, where 
input and output of the Turing machines are encoded in binary.
\end{theorem}
{\bf Proof.} We first show how to simulate a Turing machine efficiently 
with the help of a three counter machine. Given a concise encoding of the input
it is well known how to do this (see, e.g., the Theorem on p.~2 of \cite{Schroeppel73}).
Thus we will only sketch this part and focus on the input- and output-problem.

Let Turing machine $M$ computing function $f(x, y)$
have the tape alphabet $\Sigma$ that includes a blank symbol, symols 0, 1 
and a separator \#. We assume that the input is encoded as $\mbox{bin}(x)\#\mbox{bin}(y)^R$
(where $w^R$ is the reversal of string $w$), $M$ starts its operation with its tape head 
before the first symbol of $\mbox{bin}(x)$, and $M$ stops with  $\mbox{bin}(f(x, y))$
as its tape contents with its head again before the first symbol
(the simulation can easily be adopted to other input-/output-conventions). 

A three counter machine $C$ simulating $M$ has $x$ on counter~1 and 
$y$ on counter~2. For encoding $M$'s tape alphabet $C$ reserves 
$k = \lceil\log_2|\Sigma|\rceil$ bits per symbol and uses the following codes for
$M$'s symbols:
\begin{description}
\item[blank:] A sequence of $k$ zeroes (this in mandatory, since the infinite number of blank symbols
is represented by counter value 0).
\item[\#:] The sequence $0^{k-1}1$.
\item[0, 1, and further symbols:] Sequences  $0^{k-2}10$,  $0^{k-2}11$ and so on.
\end{description}
 
First $C$ computes $2^{k}x+1$ (thus encoding a separator) 
and then (using counter~3 as scratch memory) repeatedly divides counter~2 by 2 and 
multiplies counter~1 by $2^{k}$ adding the appropriate constants encoding 0 and 1.
This process continues until counter~2 is 0. 

In the next loop $C$ decodes $k$ bits from counter counter~1 and puts the encoding on 
counter~2 until the encoding of separator \# has been transferred. Notice that the least significant
bits of $y$ are encoded in the least significant $k$-bit blocks.

Finally the analogous process is carried out for $x$, putting the blocks onto counter~2. 

After this preparation, $C$ carries out the standard simulation of $M$ treating blocks of 
$k$ bits as an encoding of symbols from $\Sigma$. 

When $M$ stops, $C$ translates the encoding back to a number stored on counter~1 by reversing
the encoding outlined above. Since
all numbers can be encoded in $O(\log x+y)$ bits, the simulation is polynomial
in the input values

For the simulation of a polynomial time $k$ counter machine by a Turing machine observe
that due to the limited arithmetic the numbers generated on the conuters are polynomial and can
be encoded in linear space. Therefore a Turing machine can simulate the counter machine
by updating $k$ binary strings representing the counter contents.  \qed

\end{document}